\documentstyle[preprint,aps]{revtex}

\begin{document}
\title{Ultrafast Collective Dynamics in the Charge-Density-Wave Conductor K$_{0.3}$%
MoO$_{3}$}
\author{Yuhang Ren$^{1}$, Zhu'an Xu$^{2}$, and Gunter L\"{u}pke$^{1}$}
\address{$^{1}$Department of Applied Science, College of William and Mary,\\
Williamsburg, VA 23187\\
$^{2}$Department of Physics, Zhejiang University, Hangzhou, Zhejiang 310027,%
\\
P. R. China}
\date{\today }
\maketitle

\begin{abstract}
Low-energy coherent charge-density wave excitations are investigated in blue
bronze (K$_{0.3}$MoO$_{3}$) and red bronze (K$_{0.33}$MoO$_{3}$) by
femtosecond pump-probe spectroscopy. A linear gapless, acoustic-like
dispersion relation is observed for the transverse phasons with a pronounced
anisotropy in K$_{0.33}$MoO$_{3}$. The amplitude mode exhibits a weak
(optic-like) dispersion relation with a frequency of 1.67 THz at 30 K. Our
results show for the first time that the time-resolved optical technique
provides momentum resolution of collective excitations in strongly
correlated electron systems.
\end{abstract}


A charge-density wave (CDW) incorporates a periodic modulation of the
crystal's valence charge and is usually accompanied by a small periodic
lattice distortion \cite{MonceauBook,Gruner}. The symmetry-breaking charge
modulation lowers the energy of the occupied electronic states and raises
that of the unoccupied states, opening up a band gap, so that the CDW state
becomes stable below a certain critical temperature $T_{CDW}$ \cite{Gruner}.
Recent investigations of CDW conductors have been concerned with the
low-energy collective excitations because of their dramatic dynamical
responses, including loading and unloading cycles, plastic motion, and
intermittent avalanches, which are directly related to the collective motion
of magnetic flux quanta in high-temperature superconductors and
charge-ordering fluctuations in colossal magnetoresistance manganites \cite
{MonceauJPF,DumasRev,DessauScience2001,KidaPRB2002,KaindlScience2000,DemsarPRL1999}%
. However, very few experimental results on collective mode spectra are
available, especially in the low-energy region \cite{MonceauJPF,DumasRev}.

Ultrafast optical spectroscopy has provided important information on the
intricate dynamics of collective modes and quasi-particles in metals \cite
{EesleyPRL1983,BeaurepairePRL1996,KoopmansPRL2000}, and more recently
transition metal oxides \cite
{KaindlScience2000,DemsarPRL1999,DodgePRL1999,KisePRL2000}. The coherent
excitation of an amplitude mode (amplitudon) has been observed by
time-resolved transient reflectivity experiments in the quasi-one
dimensional CDW conductor, K$_{0.3}$MoO$_{3}$, and the photo-generation
mechanism has been described by displacive excitation of coherent phonons
(DECP) \cite{DemsarPRL1999}. However, to our knowledge the photo-induced
excitation of the low-energy phase modes (phasons), which are acoustic-like
and therefore govern the low-frequency CDW dynamics \cite{MonceauJPF}, have
not yet been reported.

Here, we report on the first time- and momentum-resolved optical
spectroscopy of collective CDW excitations in the quasi-one dimensional
conductor, K$_{x}$MoO$_{3}$ (KMO) with $x=0.3$ and $0.33$, in the
long-wavelength limit below the CDW transition temperature, $T_{CDW}$. We
present the dispersion relation and anisotropy of the transverse phase mode
in KMO in the low-frequency range from 5 -- 40 GHz. The transverse phason
exhibits an acoustic-like linear dispersion relation with a phase velocity
of $c_{0}=2.4\pm 0.3\times \ 10^{3}$ m/s along the chain direction. In red
bronze, K$_{0.33}$MoO$_{3}$, the phason oscillations reveal a pronounced
anisotropy and vanish when the probe beam polarization is perpendicular to
the chain direction. This strong anisotropy of K$_{0.33}$MoO$_{3}$
represents a clear signature of the transverse phason and is explained by
short-range correlations perpendicular to the chain direction. The amplitude
mode exhibits a weak (optic-like) dispersion relation with a frequency $%
\Omega _{+}(q=0)=1.67$ THz at 30 K. Our results show that {\it the
time-resolved optical technique provides momentum-resolved spectroscopic
information }on the low-energy dynamics of collective excitations in
strongly correlated electron systems. This important new information is
relevant for the elucidation of collective transport phenomena in high-$T_{C}
$ superconductors and colossal magneto-resistance materials.

In the time-resolved optical experiments, a KMO single crystal is excited by
150-fs pump pulses at 1.55-eV photon energy delivered by a Ti:sapphire
regenerative amplifier operating at 1-KHz repetition rate. The KMO single
crystals (5 mm$\ast $5 mm$\ast $0.5 mm) are $c$-axis oriented. The probe
beam, polarized parallel to the sample surface, probes directly the ($%
a^{\prime }-b$)-plane reflectivity, where $a^{\prime }$ represents the [102]
direction of the KMO single crystal. The $b$-axis is the direction along the
chains. An optical parametric amplifier (OPA-800C, Spectra-Physics) provides
the 150-fs probe pulses tunable in wavelength from 400 nm to 10 $\mu $m. The
unfocused pump beam, spot-diameter 2 mm, and the time-delayed probe beam are
overlapped on the sample with their polarization perpendicular to each
other. The typical pump beam power is less than 4 mW, and the probe beam
power is less than 1 mW. The change of the reflected probe beam intensity ($%
\Delta R$) induced by the pump beam is recorded as a function of time delay
for different wavelengths of the probe beam ($\lambda _{probe}=400$ nm - 2.5 
$\mu $m). A SR250 gated integrator \& boxcar averager, and a lock-in
amplifier are used to measure the transient reflectivity change $\Delta R$
of the probe beam.

Figure 1 shows the time evolution of $\Delta R$ for a K$_{0.3}$MoO$_{3}$
single crystal at 35 K. The data are taken with pump and probe wavelength at
800 nm. The probe beam polarization is parallel to the chain direction ($%
E\parallel b$). The decay of $\Delta R$ shows two damped oscillatory
components on top of a bi-exponential decay. The fast oscillations of $%
\Delta R$ are shown on a picosecond time scale in the inset of Fig. 1. The
trace shows several damped oscillations. The frequency of these oscillations
is $\sim $1.67 THz obtained by Fourier transformation. This value is in good
agreement with the frequency of the amplitude mode in K$_{0.3}$MoO$_{3}$
obtained from previous time-resolved optical measurements \cite
{DemsarPRL1999}, as well as neutron and Raman scattering \cite
{TravagliniSSC1983,PougetPRB1991}. The amplitude of the fast oscillation is
independent of the probe pulse polarization and rather isotropic in the $%
a^{\prime }-b$ plane, which is consistent with the $A_{1}$ symmetry of the
amplitude mode. Besides the fast oscillations, the data also reveal a slow
strongly overdamped modulation of $\Delta R$ (Fig. 1). The frequency of this
mode is $0.015$ THz.

In order to identify the nature of these slow oscillations we performed
spectroscopic measurements of $\Delta R$ as a function of probe wavelength
in the range from 400 to 2500 nm. Figure 2 shows the dispersion relations of
the slow and fast oscillations of $\Delta R$ in K$_{0.3}$MoO$_{3}$ at 30 K.
The wave number is given by $q=2n/\lambda _{probe}$, where $n\simeq 3$ is
the refractive index of K$_{0.3}$MoO$_{3}$ \cite{Gorshunov}. The
dispersionless amplitude mode exhibits a gap for $q\rightarrow 0$, with a
frequency $\Omega _{+}(q=0)=$ $1.67$ THz (Fig. 2). The frequency dependence
of the slow oscillation shows clearly a linear, acoustic-like dispersion.
The linear dispersion relation together with the very low frequency, $%
\upsilon =5-40$ GHz and small phase velocity, $c_{0}=2.4\pm 0.3\times \
10^{3}$ m/s, indicates the photo-excitation of a transverse phason \cite
{CLAP}. The agreement is excellent between our experimental data and the
transverse phase velocity normal to the surface, $23\pm 4$ THz\AA , obtained
by neutron inelastic scattering \cite{Hannion92PRL}.

A clear signature of the transverse phason dynamics is obtained from the
anisotropy and doping concentration dependence of period and amplitude of
the low-frequency oscillations in KMO. Figure 3 shows the time evolution of $%
\Delta R$ for probe beam polarizations parallel ($E\parallel b$) and
perpendicular ($E\perp b$) to the chain direction in K$_{0.3}$MoO$_{3}$ at
30 K (Fig. 3(a)) and K$_{0.33}$MoO$_{3}$ at 290 K (Fig. 3(b)). The data are
taken with the pump laser wavelength at 800 nm, whereas the probe wavelength
is 400 nm. The most striking result is the pronounced anisotropy of the
coherent oscillations in red bronze (Fig. 3 (b)). A strong oscillatory
signal occurs when the polarization of the probe beam is parallel to the
chain direction, $E\parallel b$. The oscillations fade away in the direction
of $E\perp b$. The amplitude of the transverse phason is expected to be
strongly anisotropic in K$_{0.33}$MoO$_{3}$, since only short-range
correlations occur along the chain direction. Coulomb interaction cannot
lead to a coherent CDW modulation on neighboring chains \cite
{TravagliniSSC1982} and therefore the coherent oscillations in $\Delta R$
disappear when the probe polarization is perpendicular to the chain
direction.

For the quasi-1D conductor K$_{0.3}$MoO$_{3}$, density wave fluctuations on
neighboring chains become correlated because of interchain interactions.
This leads to a transition to a ground state with three-dimensional,
long-range order at $T_{CDW}$ = 183 K. Coulomb interaction between
neighboring chains tends to align the chains with a certain coherence length 
\cite{GiraultPRB1989}. The coupling is different (anisotropic) parallel and
perpendicular to the chain \cite{GiraultPRB1989}. The anisotropy can be
observed from the 400-nm data of K$_{0.3}$MoO$_{3}$ (Fig. 3(a), inset) \cite
{Optic}. As the probe beam polarization is rotated from parallel to
perpendicular to the chain direction, the modulation period changes from
23.5 ps ($E\parallel b$) to 33 ps ($E\perp b$).

Since the optical properties in the two bronzes are very similar at photon
energies in the near infrared to visible range \cite{TravagliniSSC1982}, the
stronger damping observed in K$_{0.3}$MoO$_{3}$ must be related to the
damping of the transverse phason. The oscillations in K$_{0.33}$MoO$_{3}$
persist for at least $600$ ps with very little damping, whereas K$_{0.3}$MoO$%
_{3}$ exhibits only five periods of the oscillation (Fig. 3). A least-square
fit of the 400-nm data in K$_{0.3}$MoO$_{3}$ gives a damping constant $\tau
_{p}\simeq 60$ ps for the transverse phason oscillation at $\Omega
_{-}=0.035 $ THz (Fig. 3 (a)). This value is quite different from the fast
damping time, $\tau _{A}\simeq 10$ ps, observed for the amplitude mode \cite
{TravagliniSSC1983} and is larger than the values reported for the phason
from far-infrared \cite{NgPRB1986} and neutron scattering measurements \cite
{PougetPRB1991}. However, the damping constant is in good agreement with the
underdamped response from ac-conductivity measurements \cite{MihalyPS1989}.
The difference may be explained by the coupling between pinned modes and
transverse phasons in distinct frequency ranges.

Next, we present a simple physical model that accounts correctly for our
observations. In our experiments, the number of photons per pulse and unit
volume absorbed in the sample is $\sim $10$^{20}$ photons/cm$^{3}$,
comparable to the charge-carrier density ($\sim $10$^{20}$-10$^{21\text{ }}$%
holes/cm$^{3}$) in KMO; hence, one expects significant electron excitation
during ultrashort pump pulse illumination. This generates coherent
collective modes: amplitudon and phason. This excitation mechanism is
clearly different from previous pump-probe transmission experiments in
weakly absorbing materials \cite{StevensScience2001,KoehlJCP2001}. In the
latter case the wave-vector of the excited modes is determined by the
phase-matching condition of the pump beam, whereas in our case the
frequencies of the collective modes are independent of the polarization and
wavelength of the pump beam. Since the penetration depth, $\xi $,\ in KMO is
very small ($\sim $100 nm), impulsive excitation of the solid on a time
scale shorter than the material's hydrodynamic response time generates a
broad spectrum of collective modes down to frequencies of order $c_{0}/\xi
\sim 5$ GHz \cite{WrightJAP1992}.

We use a tunable optical probe pulse to detect the various frequency
components of the photo-generated collective modes. In the back-scattering
geometry (Fig. 4), part of the probe pulse is reflected by the wave front of
the excited CDW modes, and the remainder at the surface of the KMO crystal.
These reflections interfere constructively or destructively depending on the
position and time of the charge density modulation. Further, the momentum
selection rule for back scattering is $q_{i}+q_{f}=q\cos \theta $, where $%
q_{i}$ and$\ q_{f}$ are the wave vectors of the incident and scattered probe
beam in the material,$\ $and $\theta $ is the probe beam incident angle ($%
\theta \sim 0^{\circ }$). Therefore, for a given probe wavelength
phase-matching occurs exclusively on a single wave vector of the collective
excitation. This process causes the probe signal to oscillate with time
delay relative to the pump\ pulse.

In principle, both the longitudinal and transverse phason should couple to
the probe pulse and induce coherent oscillations in the time-resolved
reflectivity change, $\Delta R$. However, since the pump spot size at the
sample is quite large, the surface area within that spot is excited in a
more or less spatially uniform manner. The pump pulse is strongly absorbed
at the surface, so that the photo-excitation has a large gradient going into
the sample from the surface, and a very small gradient across the pump spot.
The wave vector component must be very nearly zero in any direction along
the surface (also along the chain direction) but can be large in the
direction normal to the surface (i.e. transverse to the chain direction).
This is similar to ultrafast excitation of acoustic wavepackets that
propagate into the sample from a strongly absorbing material \cite
{ThomsenPRB1986,HaoPRL2000}. Thus, the charge density wave propagation
develops in KMO normal to the surface, i.e., ($a^{\prime }-b$) plane, so
that only the transverse phason is generated. The novelty of this experiment
is that {\it the time-resolved optical technique can be used in a
momentum-resolved way}.

Finally, we would\ like\ to mention possible applications of our results.
The manipulation of amplitude, phase and momentum of a charge density wave
using ultrafast laser pulses represents an approach similar to coherent
control of optical phonons performed on GaAs and GaAs/AlAs superlattices, Bi
films, and crystalline quartz \cite{HaoPRL2000}. The phason and amplitudon
can be set and detected to a high degree of accuracy. In effect, the
collective mode excitation can be used as a switchable THz optical modulator
whose amplitude and phase can be controlled by ultrashort laser pulses. The
potential advantages of our variant are the magnitude of the effect,
frequency tunability, perpendicular propagation, and large lateral
dimensions.

We thank Profs. Serguei Artemenko and Keith Nelson for valuable comments and
discussions. This work is supported in part by NSF-DMR-0137322 (CWM) and the
National Natural Science Foundation of China (Grant no. 10225417).

\begin{center}
\bigskip

\bigskip

Figure Captions:
\end{center}

Fig. 1 Reflectivity change $\Delta R$ at 800-nm probe wavelength from K$%
_{0.3}$MoO$_{3}$ single crystal at 35 K. The inset depicts the fast
oscillations on a short time scale.

Fig. 2 Phason ($\Omega _{-}$) and amplitudon ($\Omega _{+}$) dispersion
relations of K$_{0.3}$MoO$_{3}$ at 30 K.

Fig. 3 Transient reflectivity change $\Delta R$ at 400-nm probe wavelength:
a) with the polarization vector $E$\ parallel ($E$ $||$ $b$) and
perpendicular ($E\perp b$) to the chain direction in K$_{0.3}$MoO$_{3}$. The
inset shows the anisotropy of the oscillation period in the $a^{\prime }-b$
plane. b) Anisotropy of $\Delta R$ for different probe polarization angles
in the $a^{\prime }-b$ plane of K$_{0.33}$MoO$_{3}$.

Fig. 4 Illustration of the photo-induced transverse phason excitation and
detection mechanism. For details see text.

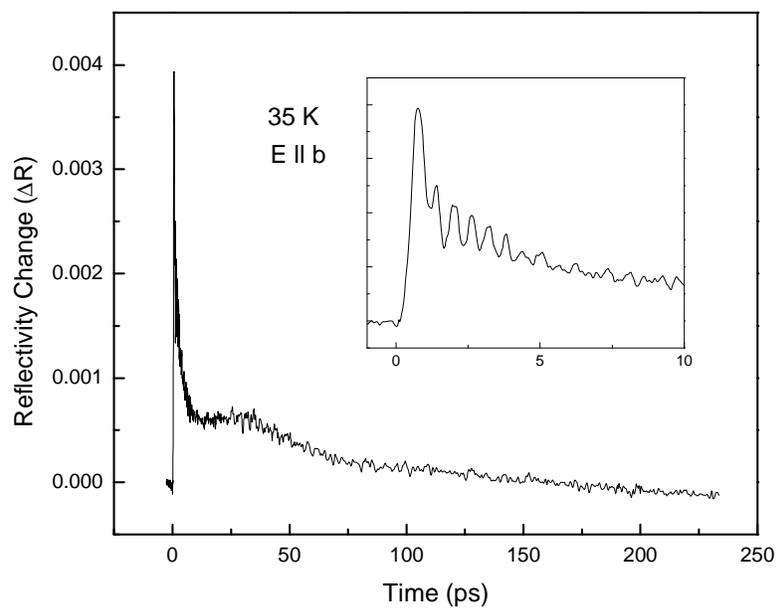
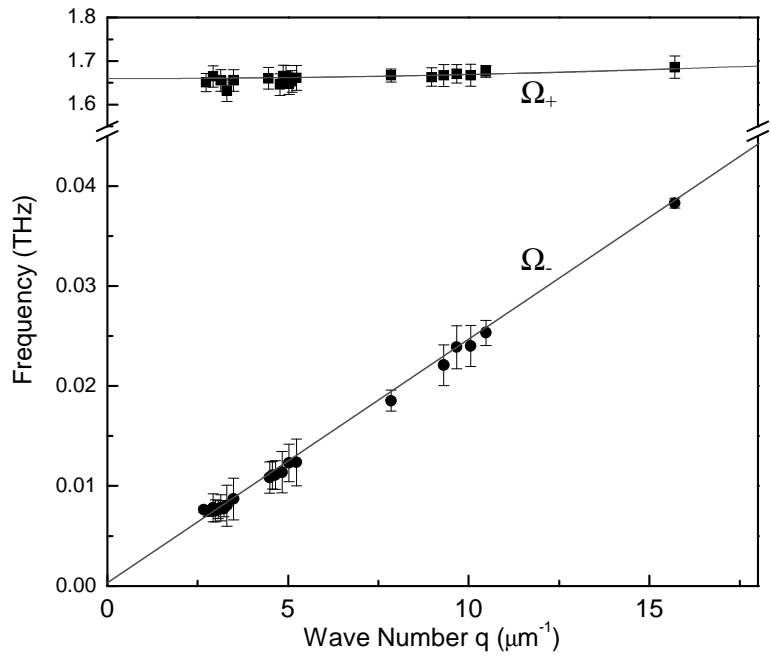
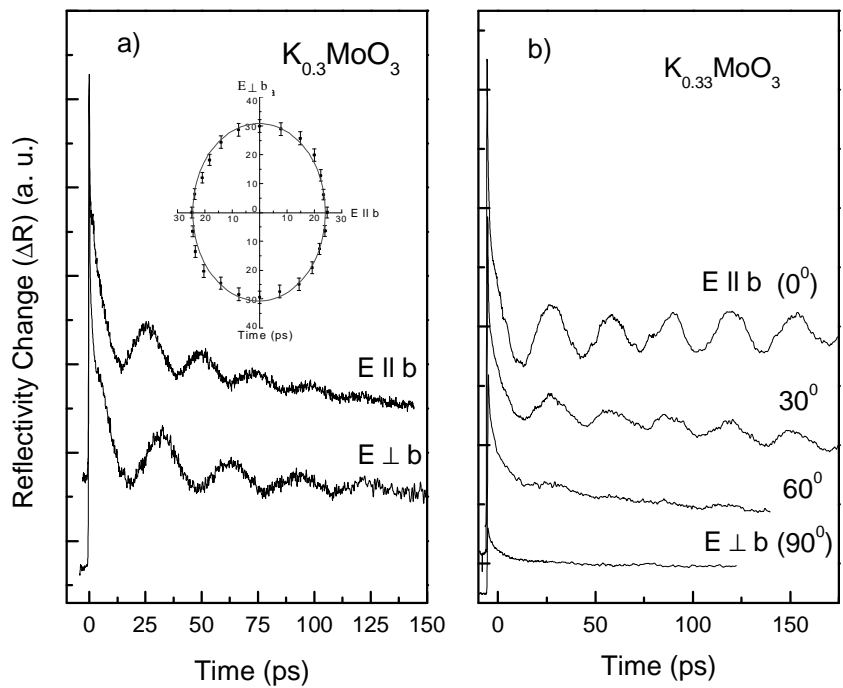
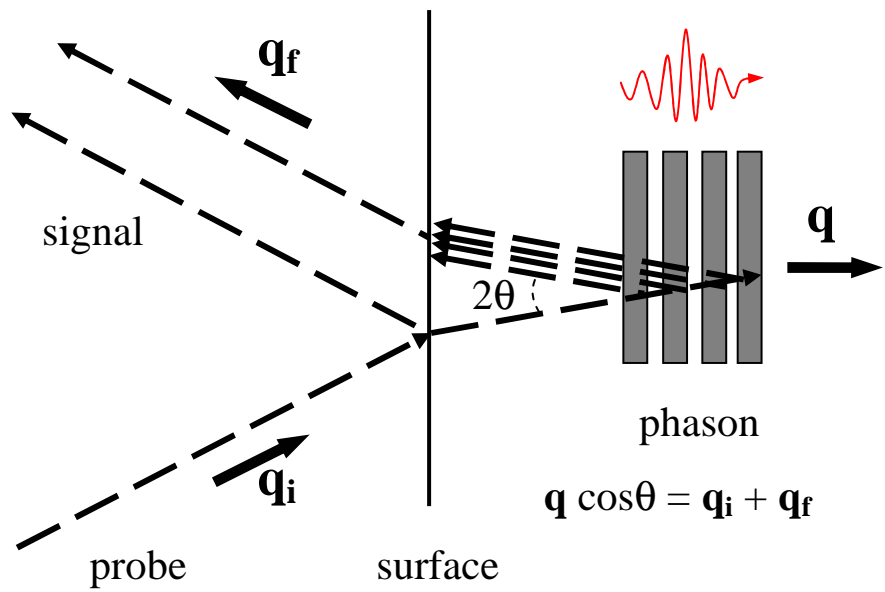
\end{document}